# Purely Speckled Intensity Images Need for SAR Despeckling with SDS-SAR

Liang Chen[1], Yifei Yin[1], Hao Shi*, Jingfei He and Wei Li

*Abstract*—Speckle noise is generated along with the SAR imaging mechanism and degrades the quality of SAR images, leading to difficult interpretation. Hence, despeckling is an indispensable step in SAR pre-processing. Fortunately, supervised learning (SL) has proven to be a progressive method for SAR image despeckling. SL methods necessitate the availability of both original SAR images and their speckle-free counterparts during training, whilst speckle-free SAR images do not exist in the real world. Even though there are several substitutes for speckle-free images, the domain gap leads to poor performance and adaptability. Self-supervision provides an approach to training without clean reference. However, most self-supervised methods impose high demands on speckle modeling or specific data, limiting their practicality in real-world applications. To address these challenges, we propose a Self-supervised Despeckling Strategy for SAR images (SDS-SAR) that relies solely on speckled intensity data for training. Firstly, the theoretical feasibility of SAR image despeckling without speckle-free images is established. A self-supervised despeckling criteria suitable for all SAR images is proposed. Subsequently, a Random-Aware sub-SAMpler with Projection correLation Estimation (RA-SAMPLE) is put forth. Mutually independent training pairs can be derived from actual SAR intensity images. Furthermore, a multi-feature loss function is introduced, consisting of a despeckling term, a regularization term, and a perception term. The performance of speckle suppression and texture preservation is well-balanced. Experiments reveal that the proposed method performs on par with supervised approaches on synthetic data and outperforms them on actual data. Furthermore, visual analysis and quantitative evaluations on actual SAR images demonstrate that our method is superior to several state-of-the-art SAR despeckling techniques.

## I. INTRODUCTION

AS a primary source of earth observation in remote sensing technology, unlike other remote sensing sensors[1], synthetic aperture radar (SAR) provides the advantage of acquiring data in all day and all weather. Therefore, SAR images have been widely used for many applications, such as target detection, marine monitoring, and geological exploration[2], [3], [4]. However, SAR images are inevitably affected by speckle noise. Speckle noise is an undesirable byproduct from the coherent summation of backscattered signals during imaging. The speckle noise degrades SAR image quality and interferes with interpretation. Consequently, speckle suppression is an indispensable step in the pre-processing of SAR images.

Numerous SAR image despeckling approaches have been proposed in the past few years. In general, the existing SAR image despeckling methods can be categorized into filter-based methods, transform domain-based methods, variational methods, Markovian model-based methods, non-local mean (NLM) methods, dictionary learning methods, and deep learning (DL) methods.

Lee filters [5], Frost filters [6], and Kuan filters [7] are the most common filter-based methods. These methods are typically assumed to operate on homogeneous regions. Thus, the edge and texture often fail to be preserved in actual SAR images. Wavelet-based [8] methods can effectively distinguish signal and noise according to the wavelet domain's characteristic difference. However, there are still some unignorable defects. The higher-dimensional features are challenging to represent, resulting in blurred texture features. Another important branch of speckle suppression is the total variation (TV)-based method [9], [10]. The TV method achieves a balance between noise suppression and edge preservation by incorporating a regularization term and a data fidelity term. While effective at noise reduction and edge texture retention, this approach inherently suffers from the limitation of compromising the overall quality of the SAR image. Markov random field (MRF) methods build the model of contextual information to provide a quantitative depiction of the prior image information [11]. Complex prior limits the application of the MRF-based despeckling methods.

By contrast, NLM methods have improved performance significantly. These methods utilize the similarities between image patches to perform weighted filtering across the entire image, which benefits the preservation of details. The methods, such as the probabilistic patch-based (PPB) filtering method [12] and block-matching 3D (BM3D) algorithm [13], exhibit satisfying performance in SAR image despeckling. However, the effectiveness of non-local mean methods depends on the setting of parameters. Moreover, it could be inefficient when processing large-scale images. As an adaptive and straightforward approach, sparse representation and dictionary learning have been able to reconstruct the image via pixel-by-pixel processing [14]. Whereas, this technique neglects the inter-pixel correlation and incurs considerable

This work was supported in National Natural Science Foundation of China No.62101041 and supported by Research Funding of Satellite Information Intelligent Processing and Application Research Laboratory. (*Liang Chen and Yifei Yin are co-first authors.*) (*Corresponding author: Hao Shi.*)

L. Chen, H. Shi, Y. Yin, and J. He are with the School of Information and Electronics, Beijing Institute of Technology, Beijing 100081, China, Beijing Institute of Technology Chongqing Innovation Center, Chongqing, 401135, China and also with the National Key Laboratory of Science and Technology on Space-Born Intelligent Information Processing, Beijing 100081, China. (e-mail: shihao@bit.edu.cn)

W. Li is with the School of Information and Electronics, Beijing Institute of Technology, Beijing 100081, China and also with the National Key Laboratory of Science and Technology on Space-Born Intelligent Information Processing.

computational cost.

In general, there are a few significant problems with most of the above methods. First, the parameters of these algorithms depend on the empirical setting. Second, the performance is scene-dependent. Third, there are sometimes artifacts in flat areas.

Recently, with the rapid development of deep learning, many denoising methods based on convolutional neural networks (CNNs) have been proposed. CNN-based methods, such as DnCNN [15], FFDNet [16] and CBDNet [17], have shown effective performance. Subsequently, Convolution Neural Networks for SAR image despeckling (SAR-CNN) [18] is proposed. SAR-CNN employs homomorphic processing to transform multiplicative speckle noise into the additive noise in the logarithmic domain. Then, it maps the despeckled image back to the original domain through the exponential function to obtain the restored image. Despite failing to achieve ideal performance, the CNN-based method has demonstrated its effectiveness for SAR image despeckling. Then, the image despeckling convolutional neural network (ID-CNN) [19] and a non-linear end-to-end mapping between the noisy and clean SAR images with a dilated residual network (SAR-DRN) [20] methods are proposed, which further improve the efficiency of speckle suppression. Subsequently, a despeckling strategy employing a pre-trained model has been proposed in [21], exhibiting commendable performance. However, the above supervised-learning based despeckling methods focus on novel network architectures, ignoring the key problem: the absence of speckle-free SAR images as ground truth in the real world.

Within the context, two approaches are mainly adopted in the literature to sidestep this problem. The first strategy is the synthetic speckle generation approach [22]. The multiplicative noise is overlayed onto optical images to simulate speckled images. The simulated speckled images are treated as inputs, while the clean grey optical images are treated as ground truth. However, the texture and edge pattern priors learned from optical images are inconsistent with the characteristics of actual SAR images. The mismatch often results in a domain gap, posing an inevitable challenge for despeckling. The second strategy is the multitemporal fusion approach [23]. A pile of actual SAR images is acquired at varying temporal instants. The clean ground truth is generated by exploiting the temporal incoherence of the stack of actual SAR images. However, collecting large multitemporal databases remains a significant challenge. On the other hand, changing the scene content may result in poor accuracy. A hybrid approach to multiply speckle simulated on SAR multi-look images is proposed to mitigate the challenges above. Nonetheless, in certain regions, the fully developed speckle hypothesis is inapplicable. The selection of correct speckle statistics remains an ongoing constraint.

In contrast, self-supervised methods provide a valid alternative since they do not concentrate on acquiring speckle-free references. Some works [31], [33], [35], [36] have begun exploring the training of networks using only speckled SAR image pairs. However, existing self-supervised despeckling methods face the following challenges. Firstly, most of these methods are still based on the Gaussian additive noise model, which is not suitable for SAR images. Moreover, they rely heavily on specific data sources, such as multi-temporal data, SLC-format data, or multi-polarization data. In fact, there is usually only intensity image data in the downstream tasks such as segmentation, object detection and so on. The high demand for specific data limits the applicability of these methods in practical scenarios. Ultimately, the despeckled images often suffer from significant degradation of texture and edge information.

In this study, a self-supervised SAR image despeckling strategy that addresses these drawbacks is proposed. Existing despeckling networks can be trained solely on actual SAR intensity images using the proposed strategy, achieving comparable despeckling performance. Experiments, including quantitative and qualitative comparisons, are conducted on both simulated and actual SAR images. Compared with the advanced despeckling methods, the results indicate that our proposed strategy can significantly suppress speckle noise and simultaneously preserve SAR image features and detailed information. The main innovations and contributions of our proposed method are summarized as follows:

1) The theoretical foundation of the self-supervised despeckling strategy for speckled SAR images has been established. The Self-supervised Despeckling Criteria for SAR images (SAR-SDC) is summarized. According to the proposed criteria, any despeckling network can be trained using only speckled image pairs.
2) The proposed RA-SAMPLE is capable of generating training pairs from a single intensity speckled SAR image, in accordance with the SAR-SDC. The generated image pairs mutually supervise each other through a cycle-consistent approach, ensuring the complete preservation and utilization of the original image information.
3) The proposed multi-feature loss function, defined as a weighted combination of three terms, takes into account the spatial and statistical properties of SAR images. Each term is responsible, respectively, for suppressing speckle, reducing spatial distortion, and preserving edges.
4) Our self-supervised approach demonstrates outstanding despeckling performance, exceeding model-based techniques, achieving comparable results to supervised learning methods training on synthetic images, and outperforming them when applied to actual SAR images. Furthermore, compared to other self-supervised methods, our approach achieves superior results.

The rest of this paper is organized as follows. Section II introduces related work, including self-supervised optical image denoising and SAR image despeckling methods. Section III illustrates the details of the proposed method. Section IV introduces the implementation details of the experiment, evaluation metrics, the compared methods, and experimental results on actual SAR images. Section V summarizes the paper.

## II. RELATED WORK

As discussed, speckle-free SAR images required for supervised methods do not exist in the real world. On this account, the self-supervised despeckling methods have developed rapidly in recent years. Self-supervised methods were initially introduced in the field of optical image denoising and later extended to SAR image despeckling.

*A. Self-supervised denoising methods in optical image*

The concept of self-supervised denoising was first proposed in Noise2Noise [24]. Noise2Noise demonstrated that the denoising network can be trained using paired noisy images rather than clean reference. Specifically, the input and target must share the same underlying ground truth, meaning that the mean of training pairs equals the underlying ground truth. The training pairs are derived from different realizations of the same scene. However, obtaining multiple scene observations with the same content is a significant obstacle, especially for SAR sensors.

This issue does not arise in single-image denoising techniques. Noise2Self [25] and Noise2Void [26] introduced the concept of blind-spot network. The noise is assumed to be spatially independent, and the statistical priors are acquired about the data distribution. Blind spots are added onto the noisy image to obtain training pairs. The network can be trained using solely individual noisy images. However, specifying the data model is challenging, particularly in actual scenarios. In addition, the absence of valuable information caused by blind spots dramatically degrades the despeckling performance. Blind2Unblind[27] proposed a global mask mapper aimed at mitigating the information loss caused by blind-spot networks. However, it still fails to overcome the aforementioned limitations. Neighbor2Neighbor [28], a self-supervised framework based on a neighbor sub-sampler, bypasses the limitations above. It aims to train denoising networks solely with noisy images available instead of modifying the structure of the despeckling network. Moreover, it is no longer necessary to characterize noise statistical models. Inspired by these works, the study of self-supervised SAR image despeckling has also been promoted.

*B. Self-supervised despeckling methods in SAR image*

Many studies have attempted to despeckle without relying on speckle-free SAR images, which are derived from optical image denoising methods. A self-supervised densely dilated CNN (BDSS-CNN) [29] for blind despeckling was designed. This method extends the Noise2Noise approach to despeckling by generating paired noisy images through overlaying multiplicative speckle noise onto optical images. Then, BDSS-CNN is trained with the generated paired noisy images. Finally, the trained despeckling network is applied to despeckle SAR images in the real world. However, the discrepancy exists between the modeled noise and the speckle in the real world. Besides, a no-reference-based SAR deep learning (NR-SAR-DL) filter [30] was proposed, which employed multi-temporal SAR images of the same scene to train the despeckling network. NR-SAR-DL can better suppress speckle noise on actual SAR images, especially for preserving point targets and radio metrics. The main limitation of NR-SAR-DL is that the change of scene content may result in poor performance. SAR2SAR [31] applies a change compensation operation to mitigate the above influence. Nevertheless, temporal diversity caused by multi-temporal data has not been solved fundamentally [32].

Speckle2void [33], a self-supervised Bayesian framework for SAR despeckling, is proposed based on the blind-spot denoising approach. To compensate for the autocorrelation process of the speckle, Speckle2void employs a whitening preprocessing and a network with a variable blind-spot size. However, the performance of Speckle2void depends on accurate statistical data modeling, which limits further improvements. Yuan et al. [34] proposed a self-supervised dense dilated convolutional neural network for blind SAR image despeckling named SSD-SAR-BS. The method uses Bernoulli sampling to generate image pairs from original speckled SAR images in the real world. The generated pairs serve as input and target to train a multiscale despeckling network. Out of inadequate consideration of SAR image characteristics, the despeckling performance can be further improved. Subsequently, MERLIN [35] provided an approach to train the network using real and imaginary parts of the single look complex (SLC) image. However, preprocessing raw data is slightly more complex. The severe speckle noise in SLC images constrains the performance upper bound. Building on this, methods such as Sublook2Sublook [36] and POL-MERLIN [37] provided some excellent solutions from the perspective of SAR data. Nevertheless, these methods impose high requirements on the original data, which limits their applicability in the real-world scenarios.

In summary, existing self-supervised methods for SAR image despeckling require enhancements in terms of computational burden, the acquisition of training data, despeckling performance and model generalization ability.

III. METHOD

In this section, a self-supervised strategy is proposed to train a despeckling network with speckled intensity images. In the first part, the feasibility of self-supervised despeckling method for SAR images is demonstrated theoretically. Subsequently, the RA-SAMPLE, which consists of a specially designed sub-sampler and a decorrelator, is introduced to generate training pairs. The sub-sampler considers the randomness and the decorrelator is designed according to the projection correlation to maintain the independence between generated pairs. Then, the overview of the proposed SDS-SAR is illustrated and the convergence is evaluated. For the last part, a multi-feature loss function is introduced to balance the despeckling effectiveness and texture preservation.

*A. Self-supervised Despeckling Criteria for SAR Images (SDC-SAR)*

Generally, in the task of image denoising, noisy image $y$ and corresponding clean image $x$ are required to train the supervised network $f_\theta$. The empirical risk minimization task is to optimize the parameter $\theta$ to meet that $f_\theta(y) \to x$. With the commonly used $L_2$ loss function, the minimization task can be described as:

$$\arg\min_\theta \mathbb{E}_x\{\mathbb{E}_{y|x} \| f_\theta(y) - x \|_2^2\} \quad (1)$$

The optimum for the $L_2$ loss is found at the arithmetic mean of the observation:

$$f_\theta(y) = E_x\{x\} \quad (2)$$

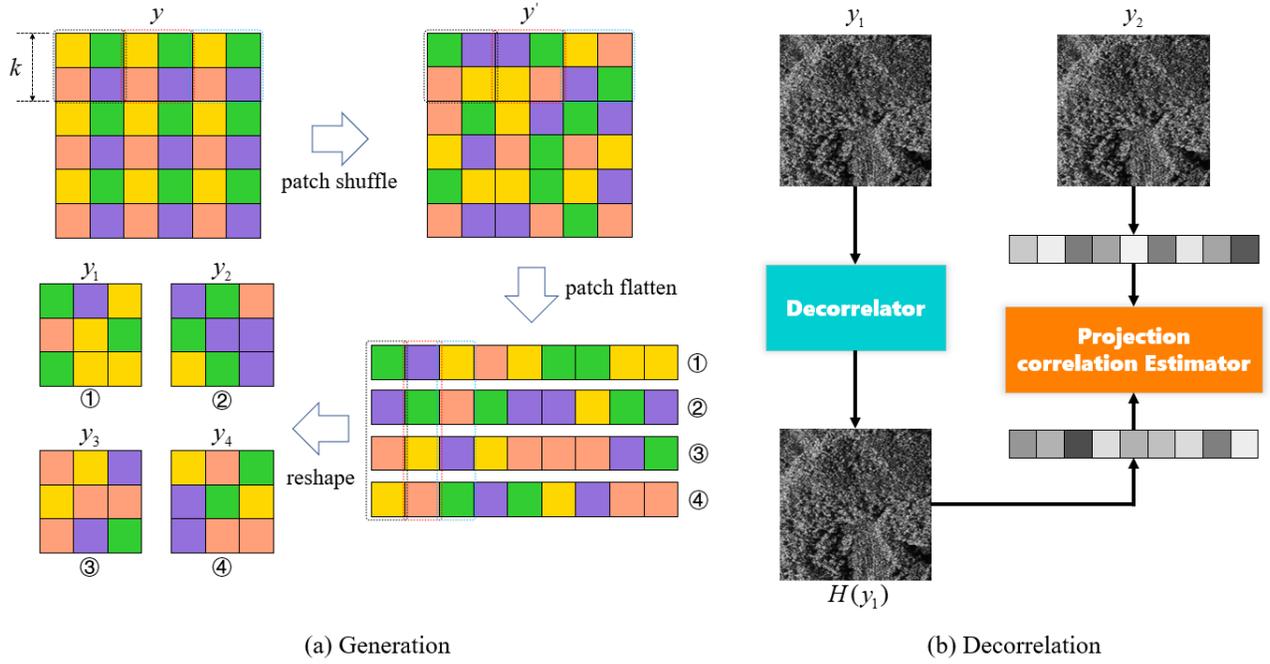

(a) Generation  (b) Decorrelation

**Fig. 1.** Illustration of the process to generate image pairs with proposed RA-SAMPLE. The generated paired images, in different colors, are generated and decorrelated in the above way. (a) The generation of image pairs. (b) The decorrelation of the generated pairs.

However, there is no speckle-free SAR image as reference in the real world. In other words, the ground truth $x$ in the equation (1) could not be found. Inspired by Noise2Noise, we consider the additive noise $z=x+n$. The minimization task can be represented as follows:

$$\arg\min_{\theta} \mathbb{E}_z\{\mathbb{E}_{y|z} \| f_\theta(y) - z \|_2^2\} \quad (3)$$

Then we transform equation (3) into a form related to $x$. There holds

$$\mathbb{E}_{y|z} \| f_\theta(y) - z \|_2^2 = \mathbb{E}_{y|x} \| f_\theta(y) - x - n \|_2^2$$
$$= \mathbb{E}_{y|x} \| f_\theta(y) - x \|_2^2 + \mathbb{E}_{z|x} \| n \|_2^2 \quad (4)$$
$$- 2\mathbb{E}_{y|x}(n \cdot (f_\theta(y) - x))$$

For convenience, the noise is assumed to be independent and zero-mean distributed, we have

$$\mathbb{E}_{z|x} \| n \|_2^2 = \mathbb{E}(n^T n) = \mathbb{E}(n) \cdot \mathbb{E}(n^T) = 0$$
$$\mathbb{E}_{y|x}(n \cdot (f_\theta(y) - x)) = \mathbb{E}(n) \cdot \mathbb{E}_{y|x}(f_\theta(y) - x) = 0 \quad (5)$$

The optimal network parameters $\theta$ of Equation (1) also remain unchanged when the targets are mutually independent and corrupted with zero-mean noise.

Speckle noise is an inherent factor leading to SAR image degradation. According to the fully-developed speckle model proposed by Goodman et al. [38], the actual SAR intensity image $Y$ can be described as:

$$Y = X \times N \quad (6)$$

where $\times$ represents the element-wise product, $X$ represents the underlying ground truth, and $N$ represents the speckle noise. Speckle noise $N$ is assumed to follow Gamma distribution with unit mean and variance $1/L$ [39]. The probability density function $P_r(\cdot)$ of $N$ can be defined as:

$$P_r(N) = \frac{L^L N^{L-1} e^{-LN}}{\Gamma(L)}, N \geq 0, L \geq 1, \quad (7)$$

where $\Gamma(\cdot)$ represents the Gamma function, and $L$ represents the number of looks. The speckle component $N$ at each pixel of the SAR image is assumed to be independent and identically distributed (i.i.d.) [40].

Equation (6) can be reformulated into the following additive noise model:

$$Y = X \times N = X + (N-1) \cdot X$$
$$= X + N' \quad (8)$$

The transformation eliminates the requirement for additional processing of the original SAR image, thereby preserving the inherent information contained within the original image. Notably, under the assumption that X and N are mutually independent, we have:

$$E\{N'\} = E\{(N-1) \cdot X\}$$
$$= (E\{N\} - 1) \cdot E\{X\} = 0 \quad (9)$$

For the single-look SAR images, the mean of $N$ is zero. Especially considering the presence of strong scatters, the pdf will become a rice distribution. The mean value of speckle remains the unit mean [48], [49]. Hence, $N'$ is zero-mean noise. In other words, any SAR image can be represented as the sum of the ground truth and zero-mean additive noise. Consequently, self-supervised despeckling of SAR images can be achieved by constructing a pair of noisy images $\{y,z\}$ containing the same ground truth:

$$E\{y\} = E\{z\} = E\{x\} \quad (10)$$

For actual SAR images, it is feasible to directly construct mutually independent noisy image pairs from the original image without requiring any transformation or modification of the original data. As long as the constructed image pairs satisfy the above conditions, they can be effectively utilized to train SAR despeckling networks. The condition is referred to as the SAR-SDC.

Further discussed, SAR2SAR [31] employs the logarithmic transformation on multi-temporal SAR images to construct speckled image pairs that satisfy the SAR-SDC condition for training. In contrast, MERLIN [35] utilizes SLC-format images and leverages the real and imaginary components to generate training pairs. In addition, image pairs satisfying the SAR-SDC condition can also be constructed using methods such as sub-band decomposition of SAR images [36], blind-spot network modeling [33], or artificially adding speckle noise [29] to the original images. These approaches adopt diverse strategies to construct training pairs, all of which conform to the SAR-SDC. The applicability of the SAR-SDC is further confirmed. Training image pairs can be generated based on SAR-SDC for coherent imaging systems affected by multiplicative noise, not limited to SAR images.

*B. Random-Aware sub-SAMpler with Projection correLation mEasurement (RA-SAMPLE)*

Acquiring paired SAR images containing the same scene can be challenging in practical applications. With respect to data sources, the multi-temporal data may introduce temporal diversity, resulting in generated image pairs contrary to the SAR-SDC. SAR images in SLC format are seldom used in practical downstream tasks, which containing more severe speckle noise compared to the SAR intensity images. Therefore, to better meet the requirements of real-world applications, this section focuses on constructing image pairs satisfying the SAR-SDC using only SAR intensity images.

The SAR intensity image is denoted as $y$ with $W$ wide and $H$ height. A sub-sampler is designed, in which a pile of noisy pairs $(y_1, y_2, \ldots, y_{k^2})$ are generated from $y$ to train despeckling network. The illustration of the process to generate image pairs with RA-SAMPLE is shown in Fig. 1, which consists of generation and decorrelation two parts.

During the generation phase, the original image is cut into small patches first. A square of side length $k$ is taken as the basic unit to divide the $y$. $[W/k] \times [H/k]$ patches can be obtained. Empirically, $k$ is set as 2. Secondly, a shuffle operation is applied on each generated patch to obtain $y'$. Then, for each patch, pixels at the same location are extracted to rearranged into a vector of length $WH/k^2$. $k^2$ vectors will be obtained after repeating the process. Finally, each vector is reshaped into a sub-sampled image of size $[W/k] \times [H/k]$.

Although random sampling operation is employed, some weak correlations may still exist between the generated sub-images. To mitigate the dependency, a decorrelation process for image pairs is introduced, as depicted in Fig.1(b). A single pair is selected from the generated image pairs to illustrate the decorrelation procedure. Initially, the image $y_1$ is processed through the decorrelator $H$, yielding the decorrelated image $H(y_1)$. Subsequently, the patch embedding process is applied to both $H(y_1)$ and $y_2$, yielding two respective vectors. Finally, the projection correlation estimator, as discussed in[41], is employed to assess the correlation between the two vectors.

Suppose that $\{(X_i, y_i), i = 1, \ldots, n\}$ is a random sample from the population $(H(y_1), y_2)$. For $i, j, k = 1, \ldots, n$,

$$a_{ijk} = I(X_i \leq X_k)I(X_j \leq X_k), b_{ijk} = ang(y_i - y_k, y_j - y_k),$$

$$\bar{a}_{i \cdot k} = n^{-1}\sum_{j=1}^{n} a_{ijk}, \bar{a}_{\cdot jk} = n^{-1}\sum_{i=1}^{n} a_{ijk}, \bar{a}_{\cdot \cdot k} = n^{-2}\sum_{i,j=1}^{n} a_{ijk},$$

$$A_{ijk} = a_{ijk} - \bar{a}_{i \cdot k} - \bar{a}_{\cdot jk} + \bar{a}_{\cdot \cdot k},$$

$$\bar{b}_{i \cdot k} = n^{-1}\sum_{j=1}^{n} b_{ijk}, \bar{b}_{\cdot jk} = n^{-1}\sum_{i=1}^{n} b_{ijk}, \bar{b}_{\cdot \cdot k} = n^{-2}\sum_{i,j=1}^{n} b_{ijk},$$

$$B_{ijk} = b_{ijk} - \bar{b}_{i \cdot k} - \bar{b}_{\cdot jk} + \bar{b}_{\cdot \cdot k}.$$

(11)

Specifically, the sample projection covariance between $H(y_1)$ and $y_2$ is defined as follows:

$$\{\text{Pcov}_n(H(y_1), y_2)\}^2 = n^{-3}\sum_{i,j,k=1}^{n} A_{ijk}B_{ijk} \quad (12)$$

The sample projection correlation between $H(y_1)$ and $y_2$, called $\text{PC}_n(H(y_1), y_2)$, is defined by

$$\{\text{PC}_n(H(y_1), y_2)\}^2 = \frac{\{\text{Pcov}_n(H(y_1), y_2)\}^2}{\sqrt{\{\text{Cvar}_n(H(y_1))\}^2 \{\text{Cvar}_n(y_2)\}^2}}$$

(13)

where $\{Cvar_n(H(y_1))\}^2 = n^{-3}\sum_{i,j,k=1}^{n} A_{ijk}^2$ and $\{Cvar_n(y_2)\}^2 = n^{-3}\sum_{i,j,k=1}^{n} B_{ijk}^2$.

If $H(y_1)$ and $y_2$ are independent, then as $n \to \infty$,

$$\frac{n\{\text{Pcov}_n(H(y_1), y_2)\}^2}{\pi/6 + S_{2n}} \quad (14)$$

Equation (14) converges to a constant governed by the standard normal random distribution.

According the result of projection correlation estimator, in practical application, a decorrelator $H$ is designed as follows:

$$H(f) = \begin{cases} \dfrac{A}{B + (f/f_c)^{2N}}, & |f| \leq f_c \\ 0 & otherwise \end{cases} \quad (15)$$

where $f_c$ is the cutoff frequency, $A$ and $B$ are empirical parameters of the model.

Theoretically, the sub-sampled images generated by RA-SAMPLE are expected to contain nearly identical ground truth. In practical applications, however, the sampling difference between the sub-sampled images needs to be evaluated. The difference between the generated sub-images resembles a noise-like image, with nearly zero mean. The detailed results will be discussed in section IV-F. Hence, for the sub-sampled images $y_1 = \text{sub}_1(y)$, $y_2 = \text{sub}_2(y)$, we have

$$y_1 = x + n'$$
$$y_2 = y_1 + diff = x + n' + n_1 \quad (16)$$

where $n'$ represents the sub-sampled noise image, i.e., $n' = \text{sub}(N)$. The mean of $n'$ remains zero. Based on the above analysis, $n_1$ is also zero-mean noise. The generated image pairs generated by RA-SAMPLE satisfy the SAR-SDC, which can be used for training.

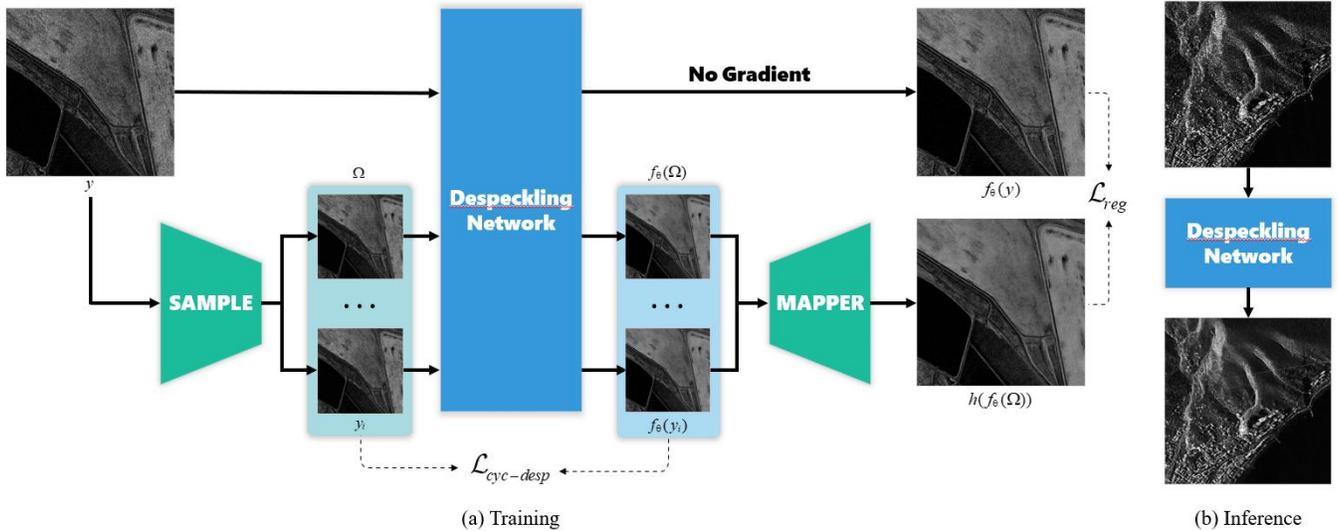

**Fig. 2.** Overview of proposed self-supervised despeckling strategy. (a) The illustration of the training phase. (b) The flowchart of the inference phase.

## C. Overview of Proposed Strategy

The overview of our proposed SDS-SAR is shown in Fig. 2. The training phase and inference phase are both illustrated.

In the training phase, a pile of speckled image pairs for training is generated with RA-SAMPLE first. Then, the generated image pairs are fed into the despeckling network for training based on the cycle-despeckling loss function, which will be detailed in Section III.E. The cyclic structure treats $y_i$ as both the input and the target, with $y_{i+1}$ serving as the corresponding target and input. Under the cycle-despeckling loss function, better despeckling performance can be achieved. Notably, the sub-sampled images contain only one-fourth of the information from the original image. Therefore, a global up-sampling mapper is designed to map the despeckled images back to their original resolution. For the generated image stack $\Omega$, the initial position of each pixel relative to the original image is recorded. According to the position information, the despeckled image pile ($f_\theta(y_1), f_\theta(y_2), \ldots, f_\theta(y_{k^2})$) is then mapped and combined into an original-size image $h(f_\theta(\Omega))$. In addition, the original SAR image $y$ is fed into the despeckling network to serve as a constraint without gradient backpropagation. The obtained $f_\theta(y)$ and $h(f_\theta(\Omega))$ are combined to form the regularization loss term.

In the inference phase, the SAR image to be despeckled is put into the well-trained despeckling network and the final despeckled image will be obtained.

## D. Convergence Analysis of the Proposed Strategy

The convergence of the proposed strategy needs to be evaluated. Concretely, in the $L_2$ norm minimization task, the expected squared difference between speckled images and corresponding ground truth can be used to demonstrate the feasibility. For clarity, let $(u_i, v_i)$ denote a single training pair. Benefiting from the sub-sampler, the content of $(u_i, v_i)$ are nearly identical, implying that $E\{u_i\} = E\{v_i\}$. Consequently, the expected squared difference between training pairs is expressed as follows:

$$\mathbb{E}_v[\frac{1}{N}\sum_{i=1}^{N}u_i - \frac{1}{N}\sum_{i=1}^{N}v_i]^2$$

$$= \frac{1}{N^2}[\mathbb{E}_v(\sum_{i=1}^{N}u_i)^2 - 2\mathbb{E}_v[(\sum_{i=1}^{N}u_i)(\sum_{i=1}^{N}v_i)] + \mathbb{E}_v(\sum_{i=1}^{N}v_i)^2]$$

$$= \frac{1}{N^2}Var(\sum_{i=1}^{N}v_i) \quad (17)$$

$$= \frac{1}{N}[\frac{1}{N}\sum_i\sum_j Cov(v_i, v_j)]$$

where $N$ is the number of training pairs. As is mentioned above, the noisy targets are not correlated with each other. The result of the above equation can be:

$$\mathbb{E}_v[\frac{1}{N}\sum_{i=1}^{N}u_i - \frac{1}{N}\sum_{i=1}^{N}v_i]^2 = \frac{1}{N}[\frac{1}{N}\sum_i Var(v_i)] \quad (18)$$

As $N$ increases, the error between noisy target for training and clean ground truth decreases. Therefore, if $N$ is large enough, the error is close to zero.

Ultimately, the empirical risk minimization task becomes:

$$\arg\min_\theta \mathbb{E}_{y_2}\{\mathbb{E}_{y_1|y_2}\{\mathcal{L}(f_\theta(y_1), y_2)\}\} \quad (19)$$

In this way, the despeckling network can be trained to obtain despeckled results without speckle-free images.

## E. Multi-feature Loss Function

In the proposed self-supervised strategy, the multi-feature loss function is designed to remove speckle noise as much as possible while preserving texture details. The characteristics of SAR images can be taken into account, and the information of the original image can be fully utilized. The proposed multi-feature loss function $\mathcal{L}$ is a weighted combination of three terms: the despeckling term, the regularization term, and the perception term.

According to the derivation in Section III-A, MSE can be used as the loss function. To enhance the upper bound of despeckling performance, it is essential to exploit all of the sub-sampled images. Furthermore, to ensure consistent supervision across these images, we design the despeckling loss function in a cycle structure, which can be expressed as:

$$\mathcal{L}_{cyc-desp} = \sum_{i=1}^{k^2} \| f_\theta(y_i) - y_{i+1} \|_2^2 \quad (20)$$

where $y_i$ is the sub-sampled noisy image and $y_{k^2+1} = y_1$. $f_\theta(\cdot)$ denotes the despeckling network.

Notably, equation (8) indicates that it is feasible to directly process multiplicative speckle noise for SAR images. This finding is also confirmed experimentally. When the training pairs are input into despeckling network without any transformation, the network is still working. Nevertheless, in practice, there might be a slight deviation $\varepsilon$ between the observation noise $N$ and the modeling noise $N_m$, as described in the following equation:

$$N = N_m + \varepsilon \quad (21)$$

If the approach of equation (8) is adopted, then there holds:

$$\begin{aligned} Y &= X \times N = X + (N_m + \varepsilon - 1) \cdot X \\ &= X + (N_m - 1)X + \varepsilon \cdot X \end{aligned} \quad (22)$$

The equation shows that the deviation $\varepsilon \cdot X$ will be added to the speckle mean. Hence, slight deviation $\varepsilon$ can be amplified through $X$, easily resulting in violation of SAR-SDC. By contrast, if the logarithmic transformation is adopted, after compensation operation of mean, we have

$$\log Y = \log(X \times N) = \log X + \log(N_m + \varepsilon) \quad (23)$$

Even if $\varepsilon$ is not zero, the impact will be suppressed due to the logarithmic damping characteristic of the log function. The experimental results confirmed that, as anticipated, the despeckling performance is improved while the training process remains more stable. To strengthen the degree of speckle suppression and accelerate the convergence of MSE [42], logarithmic operation is introduced, and the loss function can be formulated as:

$$\begin{aligned} \mathcal{L}_{cyc-desp} &= -\sum_{i=1}^{k^2} \log p(y_i \mid f_\theta(y_{i+1})) \\ &\propto \sum_{i=1}^{k^2} \frac{\log \| f_\theta(y_i) - y_{i+1} \|_2^2}{\log \| y_i - y_{i+1} \|_2^2} \end{aligned} \quad (24)$$

Considering that directly applying the despeckling term will lead to over-smoothing, a regularization term should be adopted. Training solely on sub-sampled images could result in excessive despeckling strength but insufficient preservation of edge and texture information. The original image $y$ is directly fed into the despeckling network without backpropagation to equip the network with enhanced global perceptual capability. The following regularized optimization term is applied to the despeckling network:

$$\mathcal{L}_{reg} = \| f_\theta(y) - h(\widehat{f_\theta}(\Omega)) \|_2^2 \quad (25)$$

where $\Omega$ is the despeckled sub-images pile, $\widehat{f_\theta}(\cdot)$ presents the despeckling network without backpropagation, and the $h(\cdot)$ presents the global mapping operation.

In order to capture the high-level features and preserve semantically meaningful details, a perceptual loss is used to generate high-quality despeckled images. Firstly, a Visual Geometry Group Network with 16 layers (VGG16) is pre-trained on actual SAR intensity images. It is used to extract the feature maps of the despeckled sub-image $y_i$ and the sub-image of despeckled image $f_\theta(y_i)$. Then, MSE is used to compare the features of $f_\theta(y_i)$ with $y_i$ to make the high-level information (content and global structure) closer. The perceptual term is expressed as:

$$\mathcal{L}_{per} = \frac{1}{CHW} \| \phi(f_\theta(y_i)) - \phi(y_i) \|_2^2, i = 1, 2 \quad (26)$$

There, $\phi$ is the VGG network and $\phi(\cdot)$ represents a feature map of shape $C \times H \times W$.

The final loss function is composed of three terms:

$$\mathcal{L} = \mathcal{L}_{clc-desp} + \alpha \mathcal{L}_{reg} + \beta \mathcal{L}_{per} \quad (27)$$

Specifically, the despeckling term is responsible for the reconstruction of the despeckled image space. It takes into account the characteristics of speckle noise as well as faster convergence. The regularization term prevents the despeckling effect from over-smoothing. The perception term is used to preserve image texture details and edge information.

## IV. EXPERIMENT

In this section, experiments are conducted to evaluate the despeckling performance of the proposed SDS-SAR. First, several widely used supervised networks are trained with our strategy. Meanwhile, they are employed as our baseline. Second, quantitative and visual comparison experiments are performed to illustrate the superiority of our proposed strategy. Third, comprehensive ablation studies are conducted to explain the effectiveness of our strategy further. Ultimately, the influence of sampling difference, sampling strategy and form of the loss function are further discussed.

*A. Dataset and Implementation Details*

To validate the effectiveness of the proposed SDS-SAR, extensive experiments are conducted on both simulated and actual data. Supervised learning methods require noise-free images as ground truth for training. Therefore, 2,100 grayscale images from the Merced Land Use dataset are used for simulation. The grayscale images serve as ground truth, and simulated gamma multiplicative noise are added to the ground truth to generate noisy-clean image pairs. The noise distribution follows Equation (7). The actual dataset consists of 50 high-resolution SAR images acquired from the TerraSAR, Sentinel-1, and RADARSAT-2 systems, covering areas such as ports, urban regions, and farmland.

The patch size of RA-SAMPLE is fixed as 2, and the $f_c$ of the decorrelator is set to 240. As for the parameters of loss functions, $\alpha$ is set as 1 and $\beta$ is set as 1 empirically. Adam optimizer is adopted, and its initial learning rate is set as 0.0003. Per 20 epochs, the learning rate is decayed by half. The batch size of training data is set as 4, and the number of epochs is set as 300. The experiments are conducted based on Python 3.8.19, PyTorch 1.4.0 and NVIDIA RTX A6000.

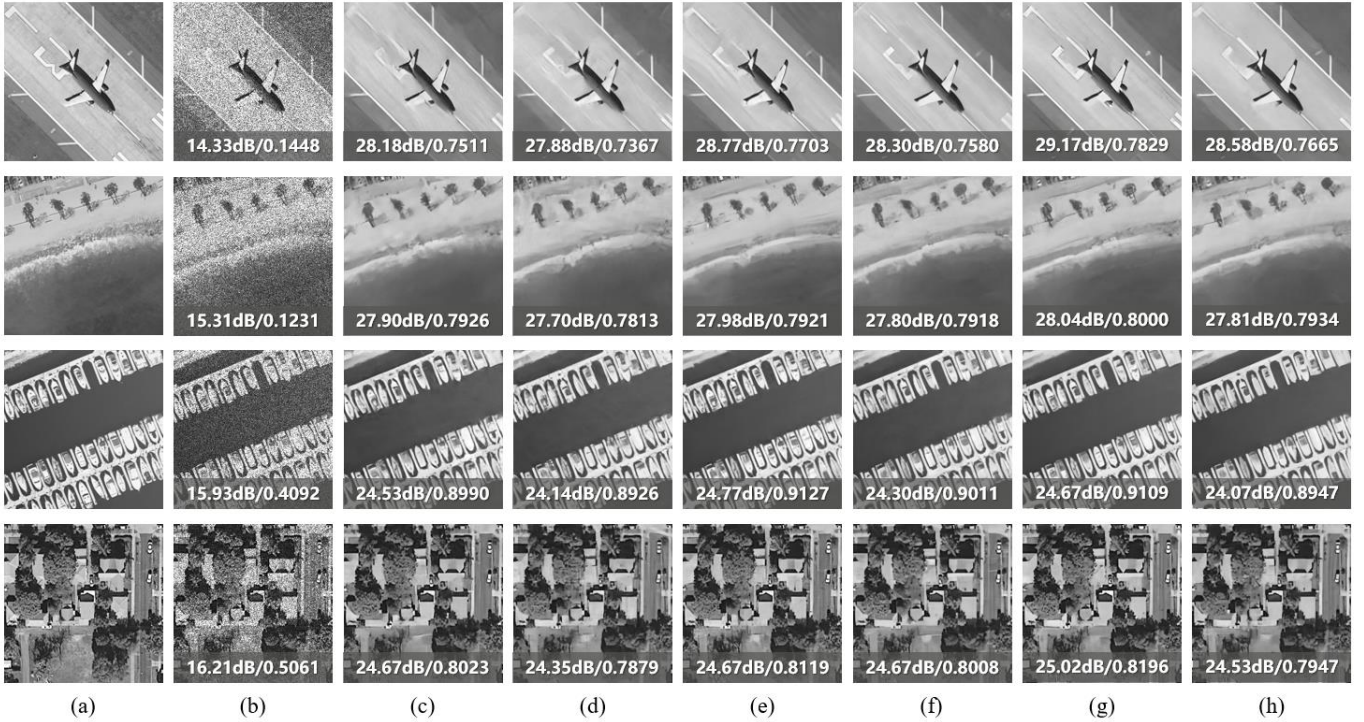

**Fig. 3.** Comparison results between supervised denoising network and the corresponding versions with SDS-SAR. (a)Clean. (b)Speckled. (c)U-Net. (d)U-Net + SDS-SAR (e)MPRNet. (f)MPRNet + SDS-SAR. (g)DeamNet. (h)DeamNet + SDS-SAR

## B. Evaluation Metrics

### 1) ENL

The equivalent number of looks (ENL) [43] is used to evaluate the performance of suppressing speckle noise by measuring smoothness in the homogeneous region. ENL is a non-referenced metric, which is defined as:

$$ENL = \frac{\mu_d^2}{\sigma_d^2} \quad (28)$$

where the $\mu_d$ is the mean of despeckled image, and $\sigma_d$ is the standard variance of despeckled image. Higher the value, better the despeckling performance.

### 2) TCR

The target-to-clutter ratio (TCR) [44] value is used to evaluate the performance of retaining the scattering information. In particular, it is essential to keep the high returns of strong scattering points in target detection tasks. TCR is to measure the difference between scattering information in the speckled and despeckled images, which can be defined as:

$$TCR = \left| 20 \cdot \log 10 \frac{\max(\hat{X})}{E(\hat{X})} - 20 \cdot \log 10 \frac{\max(X)}{E(X)} \right| \quad (29)$$

where $X$ represents the patches of speckled image and $\hat{X}$ represents the patches of despeckled images, respectively. The smaller the TCR value, the better the performance of retaining the scattering information.

### 3) MoR

The mean of ratio (MoR) [45] value is used to measure the performance of preserving radiometric information in the despeckled images. MoR is defined as:

$$MoR = \frac{1}{WH} \sum_{w=1}^{W} \sum_{h=1}^{H} \frac{X_{w,h}^{HR}}{\hat{X}_{w,h}^{HR}} \quad (30)$$

where $X$ represents the patches of speckled image and $\hat{X}$ represents the patches of despeckled images, respectively. The highest value of MoR is 1. The closer the MoR value is to 1, the better the performance of preserving radiometric information.

### 4) EPD-ROA

The edge-preservation degree based on ROA (EPD-ROA) [46] can evaluate the performance of retaining edges for a filter, which is given by:

$$EPD-ROA = \frac{\sum_{i=1}^{M} \left| \hat{X}_A(i) / \hat{X}_B(i) \right|}{\sum_{i=1}^{M} \left| X_A(i) / X_B(i) \right|} \quad (31)$$

where $\hat{X}_A(i)$ and $\hat{X}_B(i)$ represent the adjacent pixel values of filtered images along the certain direction, respectively. $X_A(i)$ and $X_B(i)$ represent the adjacent pixel values of initial images along the certain direction, respectively. If the EPD-ROA value is higher, the performance of retaining edges is better.

## C. Comparison with Baselines

### 1) Synthetic Speckled Images

For the simulation, several popular supervised desnoising networks, such as U-Net, MPRNet, and DeamNet, are used as baseline. The size of training data is set to 256 × 256. The evaluation reference metrics are Peak Signal to Noise Ratio (PSNR) and Structure Similarity Index Measure (SSIM). The supervised methods are trained based on the noisy-clean pairs. By contrast, the corresponding versions with SDS-SAR are trained directly on noisy images, without clean images as references.

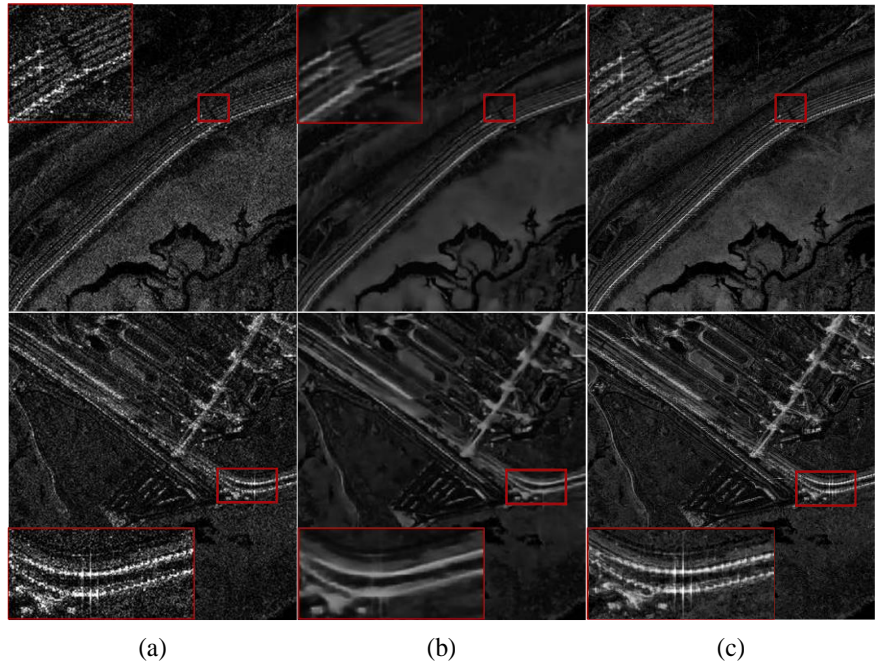

**Fig. 4.** Comparison results between SAR-CNN and SAR-CNN+SDS-SAR. (a) Speckled image. (b)Despeckling result of SAR-CNN. (c) Despeckling Result of SAR-CNN+ SDS-SAR

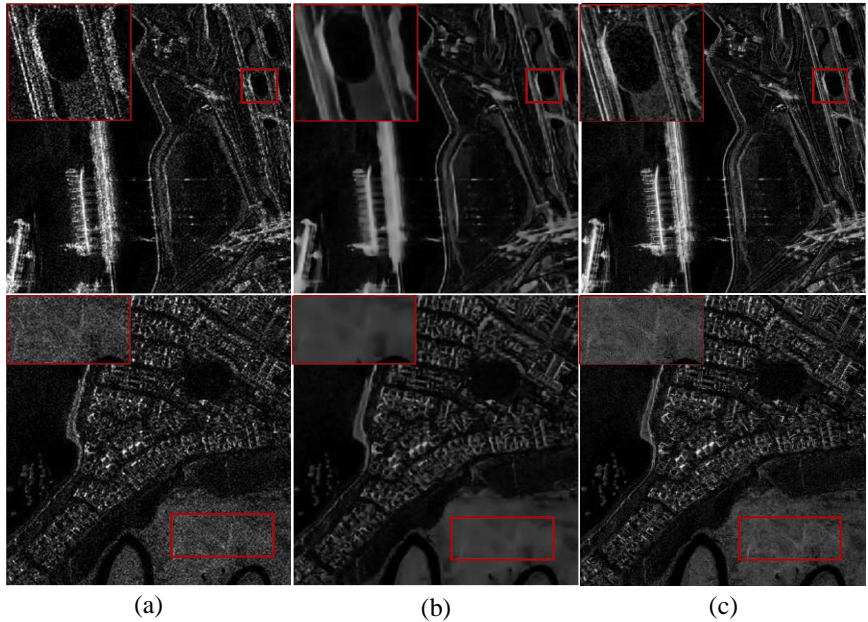

**Fig. 5.** Comparison results between SAR-DRN and SAR-DRN+ SDS-SAR. (a) Speckled image. (b) Despeckling result of SAR-DRN. (c) Despeckling result of SAR-DRN+SDS-SAR

Four typical scenes are selected for visual assessment and metric analysis, with the results shown in Fig. 3. Although clean images are not used as references, the despeckling performance of mainstream networks shows only a slight decline when employing the proposed strategy. Moreover, the method demonstrated comparable effectiveness in preserving details and textures. These results demonstrate the feasibility of training denoising networks without clean images for supervision. Furthermore, the results indicate that, with SDS-SAR, existing denoising networks can be adapted for self-supervised training.

*2) Actual SAR Intensity Images*

For the actual SAR intensity images, the SAR-CNN and SAR-DRN are employed as our baseline. Due to the absence of clean SAR images as references in the real world, the supervised networks are trained with simulated SAR images and tested on actual SAR images. Our self-supervised networks are trained directly on actual SAR images. The comparison results are presented in Fig. 4 and Fig. 5.

The synthetic SAR images are generated by adding simulated speckle onto gray optical images. However, the characteristics of SAR images are different from optical images, with scattering properties. The performance of supervised networks is not good enough due to the domain gap. As shown in the red box of Fig. 4 and Fig. 5, the texture information is blurred or lost. In addition, scattering information from some strong point targets is weakened, and the results exhibit a washing-despeckled effect.

Conversely, our strategy solves the domain gap problems and retains edges well. The overall despeckled image demonstrates that our proposed method outperforms both SAR-CNN and SAR-DRN in terms of despeckling performance. Notably, our approach exhibits significant advantages in texture preservation. Additionally, our strategy effectively retains high returns from strong point targets, ensuring that scattering intensity is sufficiently preserved.

In summary, the outstanding despeckling network trained with SDS-SAR will retain its original performance. In addition, an approach to training directly on speckled images is provided, enabling existing supervised networks to be effectively applied in real-world scenarios.

### D. Comparison with State-of-the-art

To reveal the despeckling performance of the proposed SDS-SAR strategy, the despeckling results of SAR-DRN+SDS-SAR with those of the SAR despeckling methods are compared: the PPB filter, the SAR-BM3D filter, the HDRANet, the MONet [47], the Neighbor2Neighbor, and the SSD-SAR-BS. Specifically, the adopted self-supervised methods are trained solely on intensity images, imposing minimal requirements on the data source. The despeckling results on different SAR images are shown in Fig. 6 to Fig. 8. The quantitative evaluation results of ENL, TCR, MoR, and EPD-ROA of the different methods are also listed in Table I.

*1) Qualitative Results*

As revealed in Figs. 6-8, the Lee filter can remove speckle noise well, but there is a noticeable block effect in the homogeneous region, leading to the blurring of structures. SAR-BM3D performs better than the Lee filter as the image contrast is higher and the structure is clearer. However, slight artifacts can still be observed in the SAR-BM3D despeckling results. The two methods based on supervised deep learning show better despeckling performance compared with the previous traditional methods. They are good at smoothing images to remove speckle noise. However, it is accompanied by the distortion of many image features. HDRANet, trained on synthetic SAR images, creates the problem of over-smoothing the heterogeneous area. MONet has better feature retention properties, but the influence of the domain gap still exists. Two self-supervised strategies, trained on the actual SAR intensity images, perform better than supervised ones. However, it should be noted that there is an inevitable loss of details. By comparison, our SDS-SAR strategy excels at preserving features in the heterogeneous area and suppressing speckle noise in the homogeneous area.

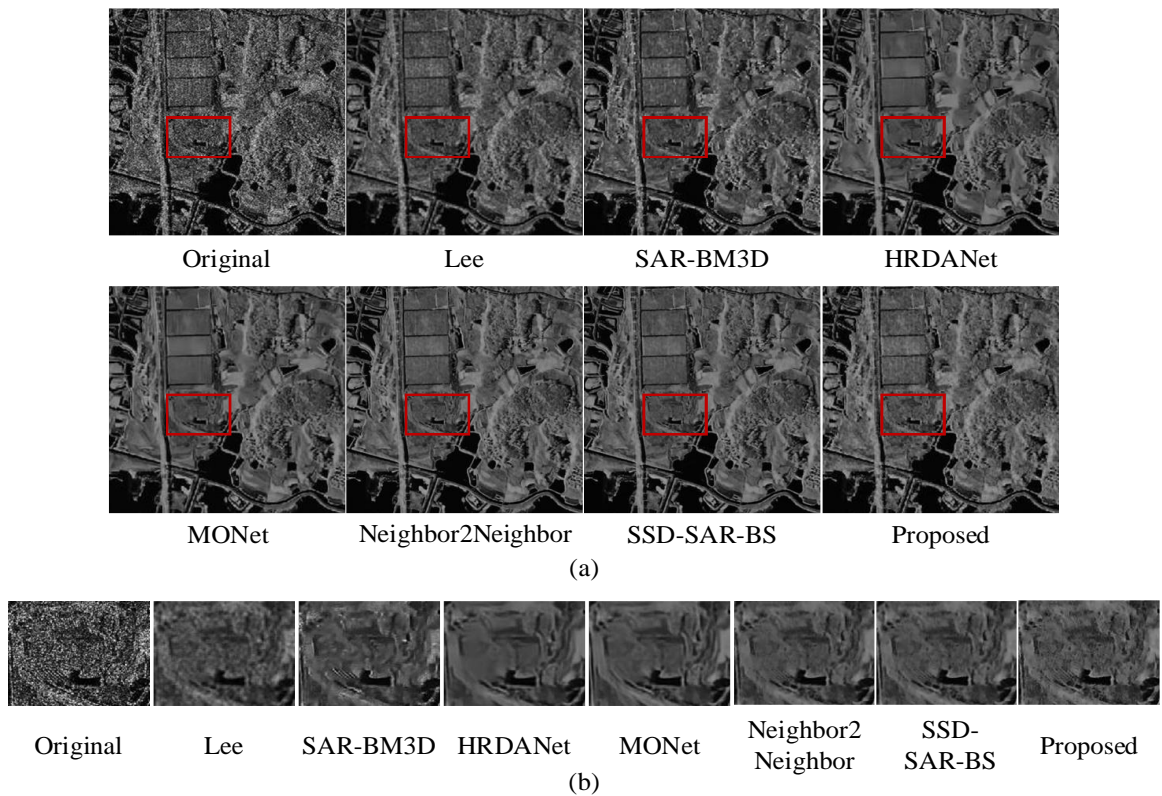

**Fig. 6.** Comparison results of our proposed SAR-DRN+SDS-SAR on Sentinel-1 image against other methods: Lee, SAR-BM3D, HDRANet, and MONet. (a) Despeckling results of different methods. (b) Details of the despeckling images.

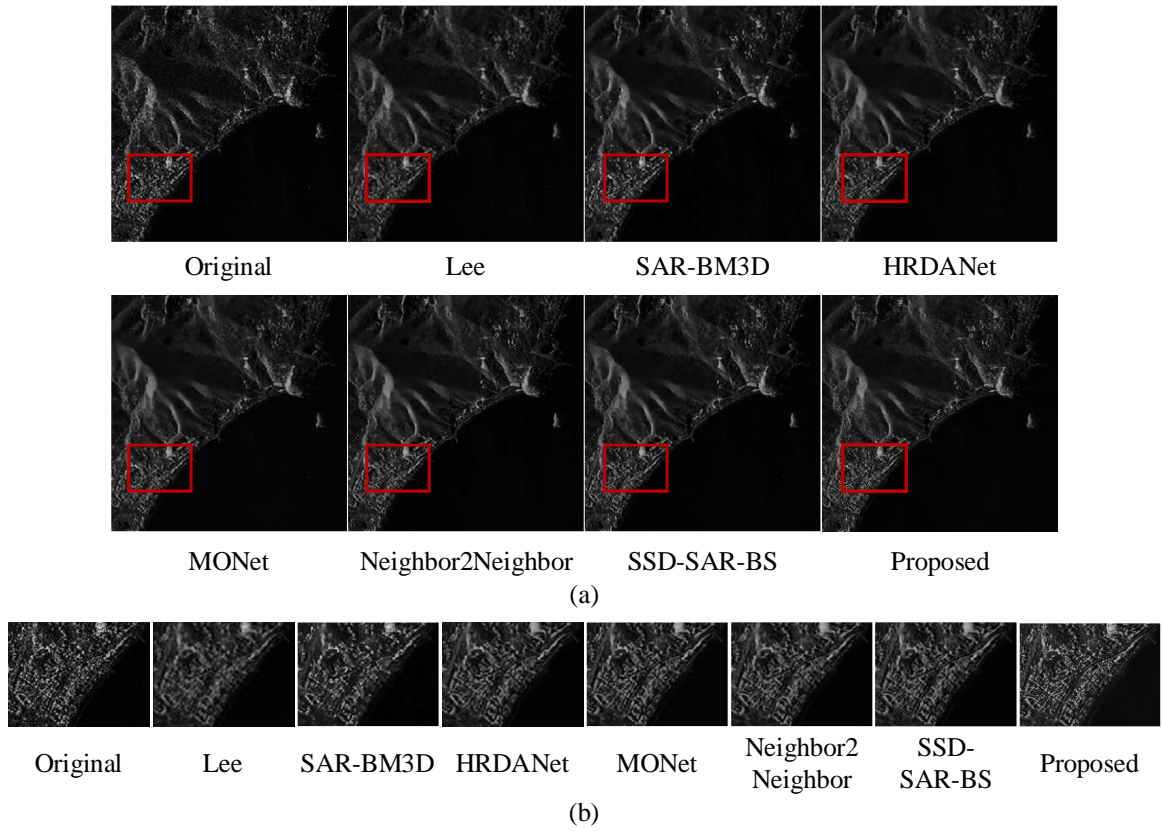

**Fig. 7.** Comparison results of our proposed SAR-DRN+SDS-SAR on TerraSAR image against other methods: Lee, SAR-BM3D, HDRANet and MONet. (a) Despeckling results of different methods. (b) Details of the despeckling images.

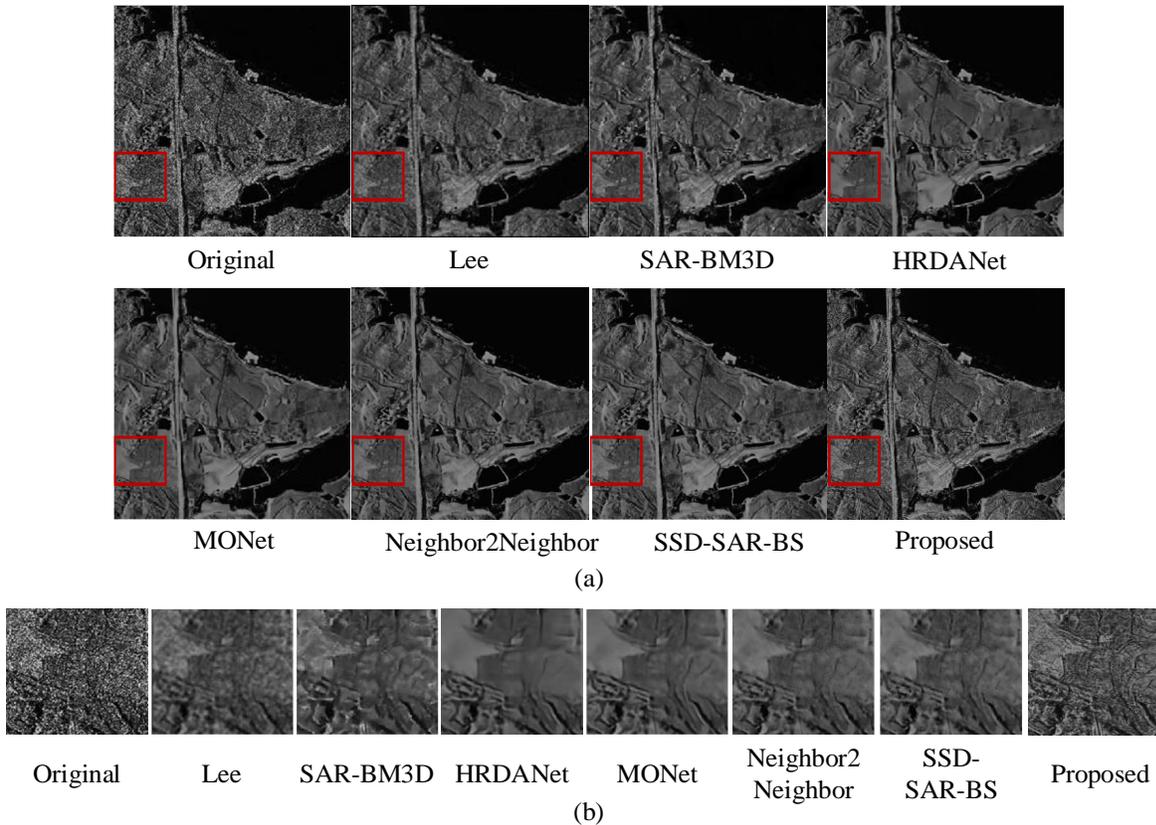

**Fig. 8.** Comparison results of our proposed SAR-DRN+SDS-SAR on TerraSAR image against other methods: Lee, SAR-BM3D, HDRANet and MONet. (a) Despeckling results of different methods. (b) Details of the despeckling images.

TABLE I
QUANTITATIVE RESULTS FOR DIFFERENT IMAGES

| No. | Methods | ENL | TCR | MoR | EPD-ROA |
|---|---|---|---|---|---|
| Image 1 | Lee | 43.8345 | 3.9190 | 0.8021 | 0.6972 |
| | SAR-BM3D | 31.2785 | 4.0905 | 0.7952 | 0.7002 |
| | HDRANet | **46.5970** | 5.0730 | 0.8120 | 0.6856 |
| | MONet | 44.2231 | 4.3958 | 0.8260 | 0.6976 |
| | Neighbor2Neighbor | 40.6252 | 3.9253 | 0.9056 | 0.7048 |
| | SSD-SAR-BS | 40.6833 | 3.9224 | 0.9083 | 0.6994 |
| | SAR-DRN+SDS-SAR(Proposed) | 40.6671 | **3.8927** | **0.9430** | **0.7064** |
| Image 2 | Lee | **5.6600** | 1.3706 | 0.7943 | 0.8441 |
| | SAR-BM3D | 3.5305 | 1.1768 | 0.6864 | 0.8497 |
| | HDRANet | 4.9029 | 1.5194 | 0.8477 | 0.8394 |
| | MONet | 4.8265 | 1.7199 | 0.9253 | 0.8414 |
| | Neighbor2Neighbor | 4.2601 | 1.2621 | 0.9212 | 0.8499 |
| | SSD-SAR-BS | 4.1952 | 1.2410 | 0.9289 | 0.8506 |
| | SAR-DRN+SDS-SAR(Proposed) | 4.3182 | **1.1193** | **0.9509** | **0.8568** |
| Image 3 | Lee | 47.9865 | 3.7501 | 0.8743 | 0.6937 |
| | SAR-BM3D | 41.9027 | 6.7534 | 0.9894 | **0.6989** |
| | HDRANet | **61.6822** | 2.5735 | 0.8524 | 0.6954 |
| | MONet | 59.9685 | 1.9055 | 0.9692 | 0.6955 |
| | Neighbor2Neighbor | 52.0348 | 1.8126 | 0.9715 | 0.6960 |
| | SSD-SAR-BS | 52.2136 | 1.7348 | 0.9729 | 0.6971 |
| | SAR-DRN+SDS-SAR(Proposed) | 52.0117 | **1.5297** | **0.9963** | 0.6980 |

To conclude, our SDS-SAR strategy can despeckle without any speckle-free reference labels. At the same time, the purpose of suppressing speckle noise in homogeneous areas while preserving the texture and edge information in heterogeneous areas is achieved. Our method achieves comparable or even superior performance to existing methods.

*2) Quantitative Results*

ENL values in Table I illustrate that the HDRANet and MONet perform well in suppressing the speckle noise. However, the presence of domain gaps leads to the issue of over-smoothing. The ENL value alone cannot fully evaluate the despeckling performance. Hence, TCR and MoR values are used to evaluate the performance of retaining the scattering information from strong point targets and radiometric preservation, respectively. Our proposed strategy, SDS-SAR, achieves the highest TCR and MoR scores.

Furthermore, SDS-SAR demonstrates a stronger ability for retaining edges as evidenced by the EPD- ROA value. Instead of using optical images to simulate SAR images, our proposed method achieves despeckling by directly learning from actual SAR intensity images. Thus, the problem of the domain gap is solved fundamentally. Specifically, according to these evaluation metrics, our proposed SDS-SAR shows an excellent ability to suppress speckle noise, preserve texture information, and preserve radiometric information.

From the perspective of visual results and assessment indices, the SDS-SAR strategy has a positive effect on all images.

*E. Ablation Study*

Here, ablation studies of our SDS-SAR strategy are conducted for further study. Specially, we assess 1) the amount of sub-sampled images pairs. 2) the importance of each term in the multi-feature loss function, 3) the influence of the weight in the multi-feature loss function.

*1)The Amount of Sub-sampled Images Pairs*

The proposed RA-SAMPLE method generates $k^2$ pairs of noisy images and employs a cycle-based supervision strategy for network training. To verify that more comprehensive supervision benefits the despeckling process, a comparison is conducted against the random-sampler method. The latter randomly selects only one pair of sub-sampled images from the original image for training. For fairness, the decorrelator is introduced into both methods. The SAR-CNN and SAR-DRN are used as the backbone for analysis on the actual SAR images from different satellites.

As shown in Table II, the proposed method achieves superior performance for the two adopted networks. It exhibits better preservation of scattering information and details, with a lower TCR and a significant improvement in the EPD-ROA metric. Although training with a single image pair is feasible, it inevitably results in significant information loss, which places a limitation on the upper bound of the despeckling performance. The employment of more image pairs can take full advantage of the training information of the original image, thereby improving despeckling performance.

*2) The Importance of Each Term in the Multi-feature Loss Function*

The three terms in the multi-feature loss function are used to suppress speckle noise, improve global awareness, and preserve texture details, respectively. Table III summarizes the performance of the SDS-SAR strategy under various combinations of these terms. First, the ENL scores are lower when the despeckling term is removed. This indicates that the despeckling results still contain a significant amount of speckle noise. Hence, the despeckling term is vital for speckle suppression. When the coefficient of the regularization term is set as 0, the corresponding despeckled image suffers from over-smoothing, leading to a significant loss of texture information. This term plays a role in global awareness, which is crucial to the texture retention. Lastly, the removal of the perceptual term decreases the EPD-ROA value. This term strengthens the concentration on the edge and the structures to improve the quality of despeckled images.

TABLE II
QUANTITATIVE COMPARISON OF RESULTS WITH DIFFERENT NUMBERS OF IMAGES PAIRS USED

| Data Source | Methods | ENL | TCR | MoR | EPD-ROA |
|---|---|---|---|---|---|
| TerraSAR | SAR-CNN + random-sampler | 74.3121 | 1.2520 | 0.9683 | 0.6788 |
|  | SAR-CNN + RA-SAMPLE | 75.4254 | 0.9611 | 0.9676 | 0.6939 |
|  | SAR-DRN+ random-sampler | 76.0337 | 1.1383 | 0.9854 | 0.6892 |
|  | SAR-DRN+ RA-SAMPLE | 76.3729 | 0.8976 | 0.9898 | 0.7102 |
| Sentinel-1 | SAR-CNN + random-sampler | 86.3344 | 2.3880 | 0.9280 | 0.7281 |
|  | SAR-CNN + RA-SAMPLE | 86.0818 | 2.1492 | 0.9338 | 0.7803 |
|  | SAR-DRN+ random-sampler | 88.2969 | 1.9503 | 0.9595 | 0.7882 |
|  | SAR-DRN+ RA-SAMPLE | 88.5568 | 1.7457 | 0.9616 | 0.8190 |

TABLE III
QUANTITATIVE RESULTS OF THE MULTI-FEATURE LOSS FUNCTION

| $L_{desp}$ | $L_{reg}$ | $L_{per}$ | ENL | TCR | MoR | EPD-ROA |
|---|---|---|---|---|---|---|
| × | √ | √ | 15.2827 | 2.9685 | **0.9151** | 0.8212 |
| √ | × | √ | 16.2351 | 3.1384 | 0.9139 | 0.8315 |
| √ | √ | × | 16.7452 | 2.7955 | 0.9038 | 0.8353 |
| √ | √ | √ | **17.3651** | **2.0540** | 0.9080 | **0.8361** |

*3) The Influence of the Weight in the Multi-feature Loss Function.*

TABLE IV
ABLATION ON DIFFERENT WEIGHTS $(\alpha, \beta)$ OF THE REGULARIZATION TERM

| ENL /EPD-ROA | $\alpha = 1$ | $\alpha = 2$ | $\alpha = 3$ | $\alpha = 4$ |
|---|---|---|---|---|
| $\beta = 1$ | **39.7188** /**0.7924** | 36.2509 /0.7870 | 38.4391 /0.7892 | 39.5712 /0.7888 |
| $\beta = 2$ | 38.3689 /0.7856 | 35.4701 /0.7892 | 35.0458 /0.7912 | 35.6777 /0.7888 |
| $\beta = 3$ | 38.4961 /0.7873 | 33.6575 /0.7873 | 39.3703 /0.7870 | 33.6945 /0.7885 |
| $\beta = 4$ | 38.5714 /0.7870 | 32.2332 /0.7867 | 36.6681 /0.7892 | 39.5062 /0.7851 |

The hyper-parameters $\alpha$ and $\beta$ are introduced in Equation (27) to control the proportion of the regularization term and perceptual term in the multi-feature loss function. Table IV lists the performance under different combinations of $\alpha$ and $\beta$ values. The ENL and EPD-ROA of the despeckled image are evaluated. Table IV shows that the weights act as a controller betwreen smoothness and details. First, a larger $\alpha$ leads to better the detail retention while the ENL of the despeckled image is reduced, which means a poor despeckling effect. Second, a larger $\beta$ leads to better details retention. However, the ENL metric tends to decrease, possibly due to the retention of speckle noise in the texture regions. In addition, the EPD-ROA reaches a peak during its increase. Based on the results, it can be empirically determined that the optimal hyper-parameters are $\alpha = 1$, $\beta = 1$.

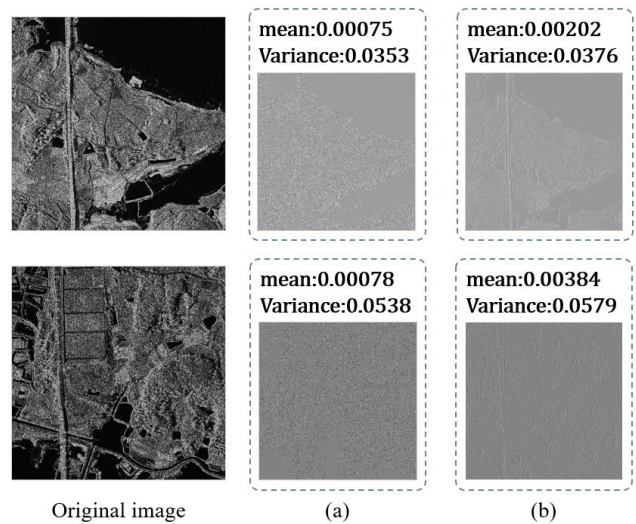

**Fig. 9.** The difference image between the sub-sampled images generated by various samplers. (a) RA-SAMPLE. (b) Sub-sampler in order.

*F. Discussion on the Sampling Difference*

In this section, a pair of sub-sampled images is randomly selected from the generated piles. Both visual inspection and mean value analysis are conducted on the difference image. As shown in Fig. 9(a), the difference between the generated sub-images resembles a noise-like image. Statistical analysis results reveals that the difference image approximates zero-mean noise, with a mean value less than $10^{-3}$.

To verify that the zero-mean noise property of the difference image is attributed to the effect of RA-SAMPLE, an additional set of experiments was conducted, and the same analysis was performed on the difference image. Specifically, pixels of the original image *y* were selected in order. The sub-sampled images and the corresponding difference image are shown in Fig. 9(b). The difference image reflects the texture information of the original image and appears as a gradient map. Mean value analysis indicates that it does not satisfy the zero-mean condition. Consequently, the generated image pairs cannot be used for despeckling network training. It can be seen that the patch shuffle in the RA-SAMPLE is of vital importance.

*G. Influence on the Sampling Strategy*

As previously mentioned, the difference between the image pairs generated using RA-SAMPLE resembles zero-mean noise images. Therefore, the generated image pairs satisfy the SAR-SDC condition and can be used for network training. In contrast, for the sub-sampler in order, the difference image between the generated sub-image pairs contain texture information from the original image. To further discussed, ENL and MOR metrics of the despeckled images are adopted to further analyze the impact of different sampling strategies. Additionally, the impact of the decorrelator is also assessed.

TABLE V
QUANTITATIVE RESULTS OF SAMPLING STRATEGY

| ENL/MoR | Figure 1 | Figure 2 |
|---|---|---|
| Sub-sampler in order | 19.6369/0.8328 | 56.0148/0.8135 |
| Sub-sampler in order with decorrelator | 19.3902/0.8509 | 51.6926/0.8278 |
| RA-SAMPLE without decorrelator | 19.9783/0.8584 | 55.3592/0.8428 |
| RA-SAMPLE | **20.3538/0.8602** | **57.2992/0.8492** |

We first substitute RA-SAMPLE with its counterpart in order and the identical decorrelator is appended respectively. The quantitative comparison is shown in Table V. The comparison displays the vital role of decorrelator in the proposed sampling strategy. Furthermore, the proposed RA-SAMPLE achieves better despeckling performance. The results further confirms that the difference between the generated noisy image pairs is essential for self-supervised SAR despeckling.

*H. Analysis on the Formulation of Despeckling Term*

In this section, we conduct experiments employing loss functions in distinct forms and analyze ENL, TCR, and EPD-ROA metrics at different epochs, as shown in Fig. 10. The despeckling results with loss function in normal form are shown in gray, while the logarithmic form results are shown in blue. For both forms, the EDP-ROA stabilizes between 0.92 and 0.94, which is not plotted for clarity.

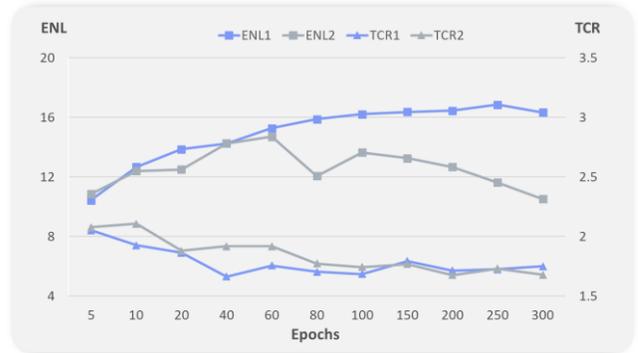

**Fig. 10.** The despeckling results of different numbers of the training epoch.

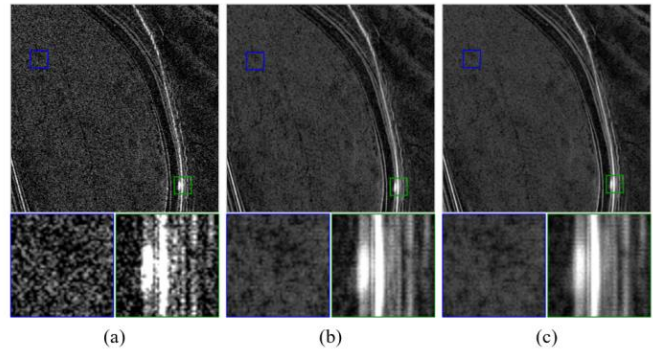

**Fig. 11.** Comparison results of the loss functions in different forms. (a) Original image. (b) Despeckling results of the normal form. (c) Despeckling results of the logarithmic form.

By contrast, the despeckling term in logarithmic form enables more stable training. Notably, as the number of training epochs increases, it suppresses speckle effectively while maintaining comparable performance in preserving texture information and strong scatterers. The results validate that the impact of the deviation ε introduced in non-ideal regions can be mitigated efficiently through adopting the despeckling term in logarithmic form. Furthermore, the despeckling results of the optimal model are shown in Fig. 11. From both visual results and metric comparisons, it is evident that the despeckling term in the logarithmic form can improve the upper bound of the despeckling performance.

## V. CONCLUSION

In this work, we have proposed a self-supervised despeckling strategy for SAR images, which can be used to the train existing supervised despeckling networks on actual SAR images. The core idea of the proposed SDS-SAR is to generate training pairs directly from individual SAR intensity images. First, SAR-SDC is introduced to guide the generation of training pairs suitable for self-supervised learning. Then, a cycle-consistent structure is employed to train the network, ensuring sufficient utilization of the original image information. In addition, a multi-feature loss function is specifically designed to balance speckle suppression and texture preservation. Experiments on both synthetic and actual

SAR images demonstrate that our proposed SDS-SAR outperforms several state-of-the-art despeckling algorithms. Furthermore, we analyze the impact of sampling difference and the formulation of the despeckling term.

Regarding the promising downstream application, with the employment of proposed SDS-SAR, the abundant existing supervised despeckling networks will be fully leveraged for actual despeckling circumstances without demanding of redundant and sophisticated speckle modeling or specific data. Therefore, SDS-SAR can be applied in broader applications since the SAR images are predominantly utilized and stored in intensity form.

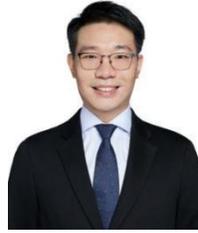

**Hao Shi** received the Ph.D degree in signal and information processing from Beijing Institute of Technology, Beijing, China, in 2016. And he worked as a post-doctor with the department of electronic, Tsinghua University. Currently, he is an associate professor at School of Information and Electronics, Beijing Institute of Technology. His research interest includes remote sensing target detection and remote sensing information real-time processing.

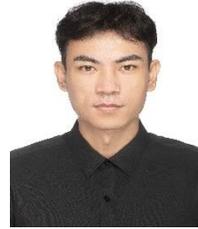

**Jingfei He** received his bachelor's degree in information engineering from Xidian University, Xi'an, China, in 2017, and the M.Sc. degree in machine learning for visual data analytics from the Queen Mary University of London (QMUL), London, U.K., in 2021. From 2017 to 2019, he served for Zhongxing Telecommunication Equipment (ZTE) Corporation, Shenzhen, China, entailing data communication and 5G. In 2023, He was a Research Assistant under the supervision of Prof. Danfeng Hong with the Aerospace Information Research Institute, Chinese Academy of Science (CAS), Beijing, China. Currently, he is a PhD student at the Beijing Institute of Technology (BIT) in Beijing, China, under the supervision of Prof. Liang Chen. His research interests include generative models and remote sensing analysis.

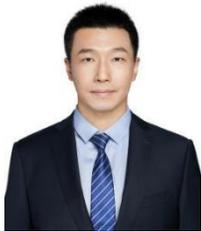

**Liang Chen** received the B.S. degree in information engineering and the Ph.D. degree in signal and information processing from the Beijing Institute of Technology, Beijing, China, in 2003 and 2008, respectively. He is currently a Professor with the School of Information and Electronics, Beijing Institute of Technology. His research interests include spaceborne real-time information processing technology and remote sensing image processing.

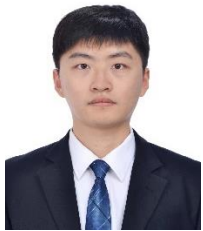

**Yifei Yin** received the B.E. degree in electronic and information engineering from Beijing Institute of Technology, Beijing, China, in 2023. He is currently pursuing the Ph.D. degree with the signal and information processing, Beijing Institute of Technology. His research interest includes SAR image intelligent interpretation and processing.

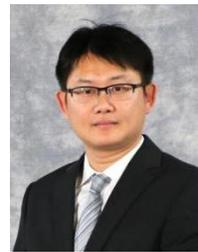

**Wei Li** received the B.E.degree in telecommunications engineering from Xidian University, Xi'an, China, in 2007, the M.S. degree in information science and technology from Sun Yat-Sen University, Guangzhou, China, in 2009, and the Ph.D. degree in electrical and computer engineering from Mississippi State University, Starkville, MS, USA, in 2012. Subsequently, he spent 1 year as a Postdoctoral Researcher at the University of California, Davis, CA, USA. He is currently a professor with the School of Information and Electronics, Beijing Institute of Technology. His research interests include hyperspectral image analysis, pattern recognition, and target detection. He is currently serving as Associate Editor for the IEEE Transactions on Geoscience and Remote Sensing (TGRS). He has served as Associate Editor for the IEEE Journal of Selected Topics in Applied Earth Observations and Remote Sensing (JSTARS) and IEEE Signal Processing Letters (SPL).